\documentclass[aps,prl,reprint,amsmath,amssymb, superscriptaddress]{revtex4-1}
\usepackage[T1]{fontenc}
\usepackage[latin1]{inputenc}
\usepackage{babel}
\usepackage{graphicx}
\usepackage[unicode=true,bookmarks=false,breaklinks=false,pdfborder={0 0 1},colorlinks=false]{hyperref}
\usepackage{color}
\usepackage{braket}
\usepackage[dvipsnames]{xcolor}

\usepackage{ulem}
\newcommand{\bs}[1]{{\boldsymbol{#1}}}
\newcommand{\bk}{\bs{k}}
\newcommand{\br}{\bs{r}}

\setcitestyle{numbers,square}
\begin{document}

\title{Tailoring  propagation of light via spin-orbit interactions in correlated disorder}

\author{Federico Carlini}
\affiliation{Laboratoire Kastler Brossel, Sorbonne Universit\'e, CNRS, ENS-PSL Research University, Coll\`ege de France, 4 Place Jussieu, 75005 Paris, France }
\affiliation{MajuLab, International Joint Research Unit UMI 3654, CNRS, Universit\'e C\^ote d'Azur, Sorbonne Universit\'e, National University of Singapore, Nanyang Technological University, Singapore}
\author{Nicolas Cherroret}
\affiliation{Laboratoire Kastler Brossel, Sorbonne Universit\'e, CNRS, ENS-PSL Research University, Coll\`ege de France, 4 Place Jussieu, 75005 Paris, France }

\begin{abstract}
Based on the fundamental interplay between spatial wavefronts and polarization degrees of freedom, spin-orbit interactions (SOI) of light constitute a novel tool for optical control at the nanoscale.   
While well described in simple geometries, SOI of light in disordered environments, where only a partial knowledge of the material's microscopy is available, remain largely unexplored.
Here, we show that in transversally random media, the disorder correlation can be exploited to tailor a variety of trajectories for ballistic beams  via SOI. In particular, we unveil the existence of an oscillating spin Hall effect, stemming from the deformation of the phase of the wavefront due to SOI. 
In  combination with a weak measurement, this phenomenon can also be maximized by an optimal choice of the disorder correlation.
\end{abstract}

\maketitle

Spin-orbit interactions of light refer to the interplay between the polarization and wavefront of optical beams, usually encoded  in spin and orbital angular momenta. This mechanism is attracting a lot of attention as it brings about a vast number of potential applications in the control of light at the nanoscale, the optical manipulation of small objects or for metrology purposes in nanostructures (see \cite{Bliokh15, Cardano15} for recent reviews). A particular manifestation of SOI of light is the optical spin Hall effect (SHE), an analogue of the electronic spin Hall effect that lies at the core of spintronics \cite{Wunderlich10, Awschalom07}.  In optics, the SHE describes transverse beam shifts occuring at a sub-wavelength scale \cite{Ling17}. Although naturally small, SHEs have been observed at interfaces using weak-measurement methods \cite{Hosten08, Qin09, Gorodetski12}, in glass cylinders exploiting multiple reflection \cite{Bliokh08} or in non-paraxial configurations \cite{Haefner09, Herrera10, Roy14}. 
An important class of systems exhibiting SHEs are inhomogeneous materials. A seminal example are gradient-index media: while geometrical optics predicts that beam trajectories are not affected by polarization, at the wave level circularly polarized beams experience helicity-dependent transverse shifts  \cite{Dooghin92, Liberman92}. Akin to the electronic SHE where the electron spin couples to a potential gradient, in optics the photon helicity couples to the refractive-index gradient,  a mechanism that can be interpreted in terms of geometric Berry phase \cite{Bliokh04, Onoda04}.

\begin{figure}
\begin{center}
\includegraphics[width=1\linewidth]{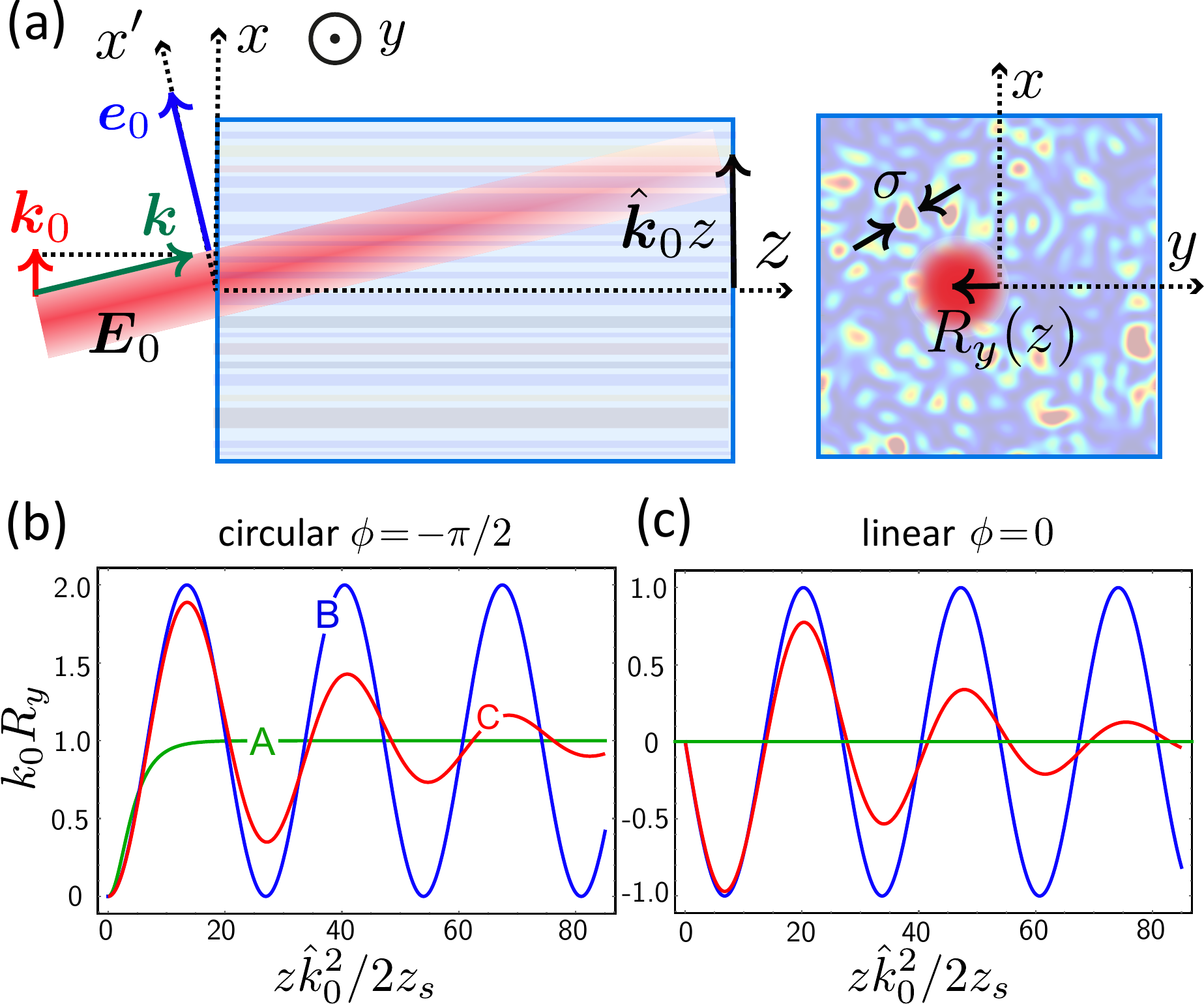}
\end{center}
 \caption{
(a) Propagation of a tilted beam of transverse wave vector $\bk_0=k_0\hat{\boldsymbol{x}}$ in a medium disordered in plane $(x,y)$ (correlation length $\sigma$). At a distance $z$, the beam centroid along $x$ is $\smash{\hat{k}_0 z}$. Due to SOI, the beam also has a transverse motion $R_y(z)$ along $y$ (spin Hall effect).
(b)  $R_y(z)$ for a circularly polarized incident light ($\phi=-\pi/2$), at $\smash{\hat{k}_0}=0.1$. The three trajectories correspond to the points A ($k_0\sigma=0.62286$), B ($k_0\sigma=0.8749$) and C ($k_0\sigma=0.84$) in Fig. \ref{zL_zS_plots}, where SOI respectively affect only the amplitude ($z_\text{L}=\infty$), only the phase ($z_\text{S}=\infty$) and both the phase and amplitude ($z_\text{L/S}$ finite) of the wavefront.
In cases B and C, a SHE oscillating around $k_0^{-1}$ appears. 
(c)  For a beam linearly polarized at $45^\circ$ in the plane $(x',y)$ ($\phi=0$), 
a SHE is also present 
but oscillates around $0$. 
}
 \label{fig-scheme}
\end{figure}

Beyond the case of a controlled inhomogeneity, it was recently shown that a SHE of light could also emerge for beams propagating in transversally disordered media \cite{Cherroret19}. 
In practice, clarifying the role of SOI in disordered environments is important for at least two reasons. First, because disordered materials are in general more the rule than the exception, in particular at the nanoscale where SOI typically operate. In addition, recent progresses in wave control or imaging have shown the great potential of using disorder as a tool rather than as a nuisance in general \cite{Vellekoop07, Popoff10, Vellekoop10, Katz12, Katz14, Gigan17}. Whether this potential could be pushed to the realm of spin-orbit physics remains an open question.
In this Letter, we take a step in that direction by theoretically showing that the combined influence of SOI on the amplitude and the phase of the optical wavefront can be exploited to tailor a variety of transverse motions for the ballistic component of light in a disordered medium, the so-called coherent mode.
This control relies on two fundamental ingredients, neglected in \cite{Cherroret19}, the disorder correlation
and the random variations of the refractive index. 
In particular, we find that a proper choice  of the disorder correlation makes it possible to realize an \textit{oscillating} SHE, see Fig. \ref{fig-scheme}, and a corresponding oscillation of the beam polarization. This phenomenon, a consequence of the interferential nature of the coherent mode, was so far not known.
Remarkably, while such an oscillating SHE is naturally small and may be hidden by multiple scattering, we show that these drawbacks can be both overcome. First, by a weak-measurement detection scheme allowing to amplify SOI of light to the macroscopic level and, second, by a proper tuning of the disorder correlation to minimize the propagation distance at which the SHE occurs.


Consider a monochromatic, polarized optical field $\smash{\boldsymbol{E}_0(\br_\perp)=E_0(\br_\perp)\boldsymbol{e}_0}$ impinging at $z=0$ on a three-dimensional material lying in the half-plane $z>0$ [$\br_\perp=(x,y)$]. 
We choose $E_0(\br_\perp)=[2/(\pi w_0^2)]^{1/2}\exp(-r_\perp^2/w_0^2+ik_0 x)$ with $k_0w_0\gg1$, which describes 
a tilted, collimated beam of waist $w_0$, as illustrated in Fig. \ref{fig-scheme}a.
The polarization vector lies in the ($x',y$) plane, perpendicular to the direction of propagation. We take it of the form $\smash{\boldsymbol{e}_0=(\hat{\boldsymbol{x}}'+e^{i\phi}\hat{\boldsymbol{y}})/\sqrt{2}}$. 
We focus on a dielectric medium with transverse spatial disorder: its permittivity $\epsilon(\br_\perp)=\bar{\epsilon}+\delta\epsilon(\br_\perp)$ has a random component $\delta\epsilon$ in the plane $(x,y)$, but is homogeneous along $z$ \cite{Schwartz07, Boguslawski17, Cherroret18}. In practice, this geometry can be realized using two-dimensional photonic lattices imprinted onto a photo-refractive crystal \cite{Schwartz07, Boguslawski17} or a glass \cite{Bellec12}.
We model the disorder fluctuations by a random function of zero mean and Gaussian correlation, $\smash{\langle\delta\epsilon(\br_\perp)\delta\epsilon(0)\rangle=\gamma/(4\pi\sigma^2)\exp(-r_\perp^2/4\sigma^2)}$, where $\gamma$ is the disorder amplitude, $\sigma$ the correlation length and the brackets refer to disorder averaging.

As the beam propagates in the medium, the components $E_j$ ($j=x,y,z$) of the electric field obey
\begin{equation}
\left[(\Delta+\omega^2\epsilon(\br_\perp)/c_0^2)\delta_{ij}\!-\!\nabla_i\nabla_j
\right]E_j(\boldsymbol{r}_\perp,z)=0,
\label{Helmholtz_eq}
\end{equation}
with $\omega$ the frequency and $c_0$ the vacuum speed of light.
In the Letter, we study the evolution of the coherent mode, of intensity distribution is $I_c(\br_\perp,z)\!=\!|\langle \boldsymbol{E}(\br_\perp,z)\rangle |^2$, after a propagation distance $z$. 
The coherent mode refers to the portion of light propagating ballistically in the medium, i.e., in the forward direction $\hat{\boldsymbol{x}}$, as opposed to light undergoing multiple scattering \cite{Sheng95}. 
To access $I_c$, we examine the Fourier components of the average field. The latter are formally given by
$\smash{\langle\tilde{\boldsymbol{E}}(\bk_\perp,z)\rangle\!=\! \exp(i k_z z-i\boldsymbol{\Sigma}z/2k_z)\tilde{\boldsymbol{E}_0}(\bk_\perp)}$, 
where $\tilde{\boldsymbol{E}}_0(\bk_\perp)=\sqrt{2\pi}w_0\exp[-(\bk_\perp-\bk_0)^2w_0^2/4]$ and  $\smash{k_z=\sqrt{k^2-k_\perp^2}}$
with $k=\sqrt{\bar{\epsilon}}\omega/c_0$ the total wavenumber. The quantity $\boldsymbol{\Sigma}$, known as the self-energy tensor \cite{Sheng95}, encodes all effects of the disorder. Its real part describes how the phase of the  wavefront evolves on average in the disorder (mean refractive index), while its imaginary part governs the attenuation of its amplitude (extinction coefficient).  An explicit calculation detailed in the Supplemental Material (SM) gives
\begin{eqnarray}
\langle\tilde{\boldsymbol{E}}(\bk_\perp,z)\rangle&=&
\tilde{{E}}_0(\bk_\perp)e^{i k_z z}
\Big[e^{-i \Sigma_{1} z/2k_z}\boldsymbol{e}_0+\nonumber\\
&&\left(e^{-i \Sigma_{2} z/2k_z}-e^{-i \Sigma_{1} z/2k_z}\right)\boldsymbol{p}(\bk_\perp)\Big].
\label{field_eq}
\end{eqnarray}
The complex numbers $\Sigma_1$ and $\Sigma_2$, evaluated below, are combinations of eigenvalues of $\boldsymbol{\Sigma}$.  
The existence of two independent self-energies stems from the fact that as soon as the incident beam is tilted, i.e. $k_0\ne0$, the statistical isotropy in the $(x,y)$ plane is broken, making momentum conservation along $z$ (due to translation invariance along that direction) the unique symmetry of the 
problem.
In Eq. (\ref{field_eq}), the first term in the right-hand side (r.h.s.) 
describes an evolution \textit{without} spatial deformation.
Its amplitude is controlled by $\Sigma_1$, which defines the mean free path along $z$, $z_s=-k_z/\Im(\Sigma_{1})$, beyond which the coherent mode is attenuated due to scattering in other directions  \cite{Sheng95, Cherroret18}.
The second term in the r.h.s. is proportional to the projection $\boldsymbol{p}(\bk_\perp)=(\hat{\boldsymbol{e}}_\perp\cdot\boldsymbol{e}_0)\hat{\boldsymbol{e}}_\perp+(\hat{\boldsymbol{z}}\cdot\boldsymbol{e}_0)\hat{\boldsymbol{z}}$ of the polarization onto the $(\hat{\boldsymbol{e}}_\perp=\bk_\perp/k_\perp,\hat{\boldsymbol{z}})$ plane. It describes a wavefront deformation coupling polarization and momentum and, as such, encodes the phenomenon of spin-orbit interaction. 
Notice that this deformation only arises when $\Sigma_1\ne \Sigma_2$.

\begin{figure}
\begin{center}
\includegraphics[width=0.99\linewidth]{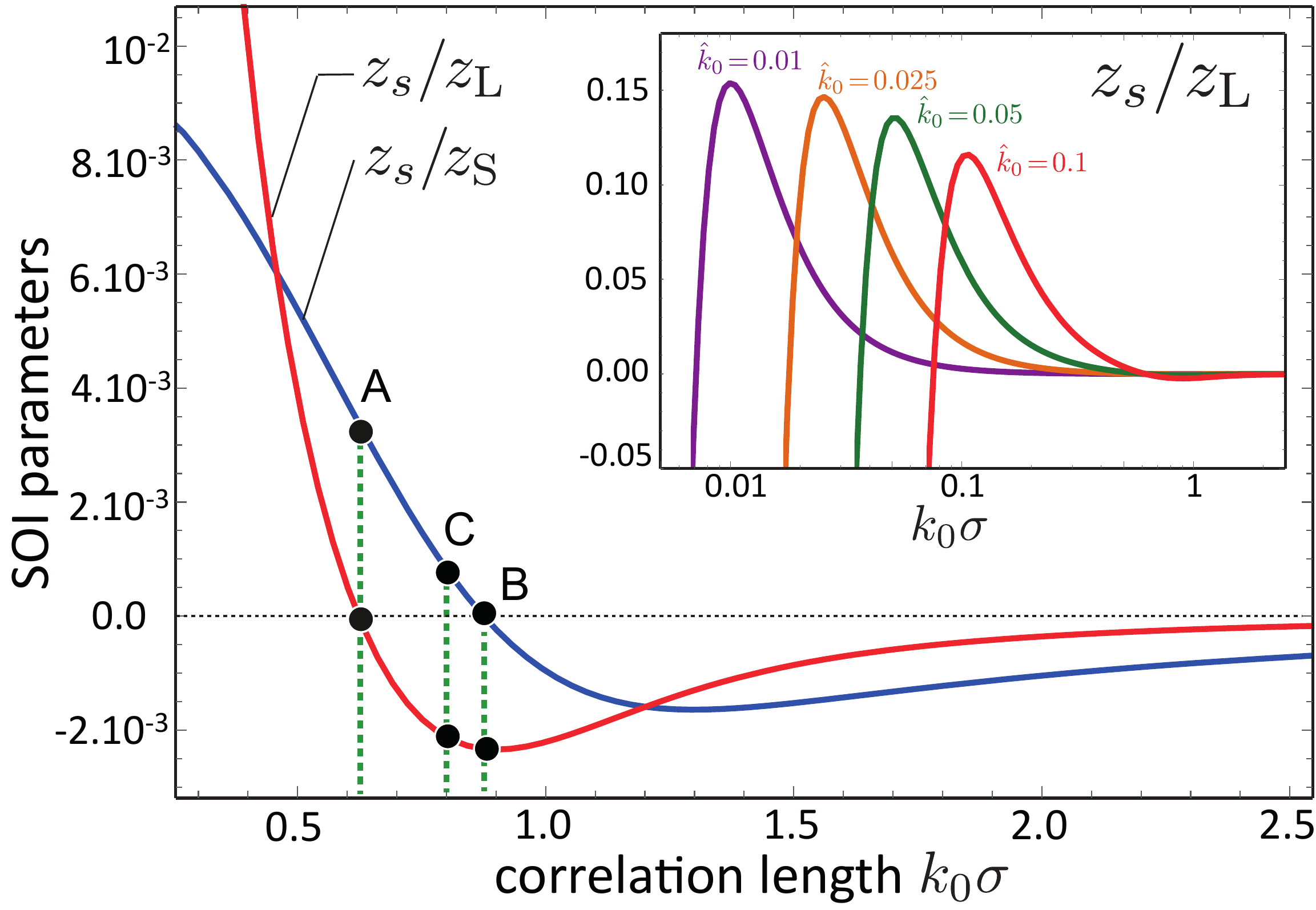}
\end{center}
 \caption{
Length scales $z_\text{L}$ and $z_\text{S}$ controlling the SHE as a function of the disorder correlation length, for $\hat{k}_0=0.1$ and in units of the mean free path $z_s$.
At points A and B, $z_\text{L}=\infty$ and $z_\text{S}=\infty$, respectively. At point C
, both $z_\text{S}$ and $z_\text{S}$ are finite. 
The inset shows that $z_s/z_\text{L}$ reaches a maximum at lower correlation lengths. At this maximum, the oscillating SHE occurs at the scale of a few $z_s$.
}
 \label{zL_zS_plots}
\end{figure}

To unveil the effect of the SOI term, we examine $\langle\boldsymbol{R}(z)\rangle
=\int d\br_\perp \boldsymbol{r}_\perp I_c(\br_\perp,z)/\int d\br_\perp I_c(\br_\perp,z)$, the beam centroid.  
The first term in the r.h.s of Eq. (\ref{field_eq}) gives the main contribution to this quantity : $\langle\boldsymbol{R}(z)\rangle\simeq \smash{\hat{\boldsymbol{k}}_0z}$. This result features a straight-line propagation with fixed polarization, see Fig. \ref{fig-scheme}a. 
It also coincides with the prediction of the paraxial limit, where $\sim\smash{\hat{k}_0=k_0/k}\ll1$. Indeed, at small angle one finds $\smash{\Sigma_1-\Sigma_2=\mathcal{O}(\hat{k}_0^2)\to 0}$, so that statistical isotropy in the $(x,y)$ is restored and the second term in Eq. (\ref{field_eq}) vanishes.
In general, however,  $\Sigma_1$ and $\Sigma_2$ differ, leading to a non-trivial component $R_y(z)$ of $\langle\boldsymbol{R}(z)\rangle$ along the \emph{transverse} direction $y$ \cite{footnote1}: 
\begin{equation}
R_y(z)\!=\!-\frac{\sin\phi}{k_0}\left[1\!-\!\frac{\cos z/2z_\text{L}}{\cosh z/2 z_\text{S}}\right]\!+\!\frac{\cos\phi}{k_0}\frac{\sin z/2z_\text{L}}{\cosh z/2z_\text{S}}.
\label{beam_shift_eq}
\end{equation}
This transverse motion is governed by two core parameters, $z_\text{S}=[\Im(\Sigma_{2}-\Sigma_{1})/k_z]^{-1}$ and $z_\text{L}=[\Re(\Sigma_{2}-\Sigma_{1})/k_z]^{-1}$, which  represent respectively the length  scales over which SOI modify the amplitude and the phase of the coherent mode. 
The first term in Eq. (\ref{beam_shift_eq}) was originally discovered in \cite{Cherroret19}, 
for an elementary model of uncorrelated disordered and discarding refractive-index effects [i.e., $\Re(\Sigma_1), \Re(\Sigma_2)=0$]. Under these approximations, $z_\text{L}\!\to\!\infty$ and $\smash{z_\text{S}\to z_s/\hat{k}_0^2}$, so that $\smash{R_y(z)=-\sin\phi/k_0[1-\cosh^{-1}(z \hat{k}_0^2/2 z_s)]}$. This describes a monotonic shift existing only for beams of finite helicity (or spin) $\sin\phi$.
In the more realistic case considered here, however, the physics pertained to Eq. (\ref{beam_shift_eq}) is much richer since the modifications of the refractive index due to SOI give rise to spatial \emph{oscillations} of the beam at the scale of $z_\text{L}$, modulating a monotonic component governed by $z_\text{S}$. In Eq. (\ref{field_eq}), these oscillations originate from the interference between  the SOI term with itself, and between the SOI term and the paraxial one.

Both $z_\text{S}$ and $z_\text{L}$ are functions of $\smash{\hat{k}_0}$, the deviation from paraxiality, and $\sigma$, the disorder correlation. We have 
computed them from $\Sigma_1$ and $\Sigma_2$, 
see SM. The results, shown in Fig. \ref{zL_zS_plots}, reveal the
 interesting feature that $z_\text{L}^{-1}$ and $z_\text{S}^{-1}$ vanish at specific values of $k_0\sigma$ (points A and B, respectively).
In practice, this offers the possibility to tailor various types of SHEs via $\sigma$.
To illustrate this idea, we show in Fig. \ref{fig-scheme}b and c the transverse motion realized for $\phi=-\pi/2$ and $0$ at points A and B, as well as at point C  where both $z_\text{S}$ and $z_\text{L}$ are finite.
In configuration A, SOI only affect the amplitude of the wavefront. 
This leads to a monotonic increase of $R_y(z)$ toward the asymptotic value $1/k_0$ for $z\gg z_\text{S}$, effectively reproducing the result of \cite{Cherroret19}. In case B, in contrast, SOI are  purely of phase origin and $R_y(z)$ exhibits oscillations around $1/k_0$. In the generic case C, finally, the oscillations are present but damped. 
Another interesting prediction of Eq. (\ref{beam_shift_eq}) is that an oscillating SHE can arise even for linearly polarized beams ($\phi=0$) i.e. without initial spin, see Fig. \ref{fig-scheme}c. We discuss this intriguing phenomenon below.

\begin{figure}
	\begin{center}
		\includegraphics[width=0.8\linewidth]{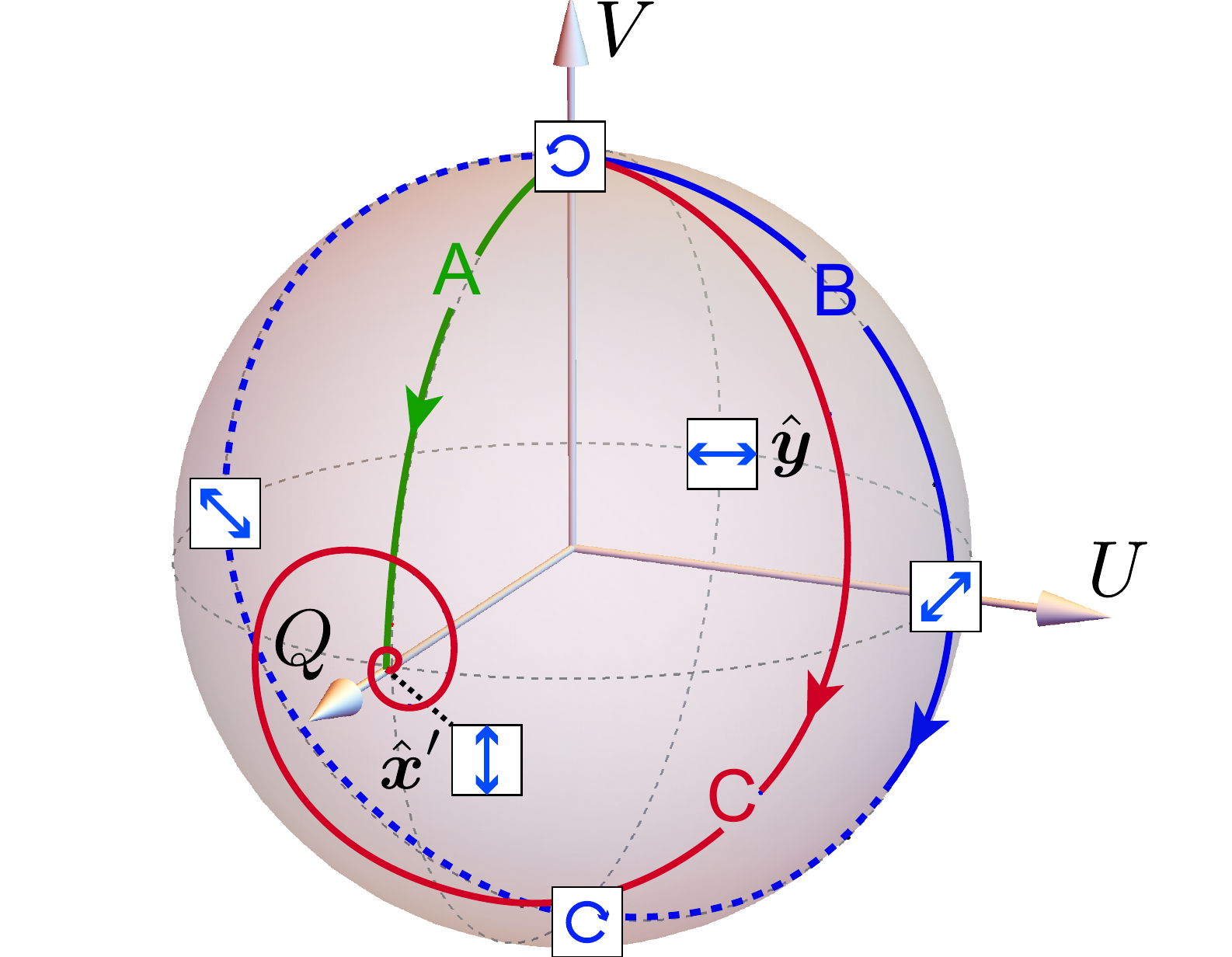}
	\end{center}
	\caption{
Evolution of the polarization state $\boldsymbol{e}(z)$ on the Poincar\'e sphere when starting from circularly polarized beam ($\phi=\pi/2$), in configurations A, B and C. Axes are parametrized by the Stokes parameters  $U$, $V$ and $Q$. 
}
	\label{fig-sphere}
\end{figure}

The SHE of the coherent mode is accompanied by an evolution of its mean polarization direction $\boldsymbol{e}(z)$. The latter follows from Eq. (\ref{field_eq}), using that the momentum distribution  of the beam always remains peaked around $\bk_0$: $\boldsymbol{e}(z)\simeq \langle\tilde{\boldsymbol{E}}(\bk_0,z)\rangle/|\langle\tilde{\boldsymbol{E}}(\bk_0,z)\rangle|$, with
\begin{equation}
\langle\tilde{\boldsymbol{E}}(\bk_0,z)\rangle
\propto\left[e^{-i \Sigma_2 z/2k_z}\hat{\boldsymbol{x}}'\!+\!e^{i\phi}e^{-i \Sigma_1 z/2k_z}\hat{\boldsymbol{y}}\right].
\label{field_eq_k0}
\end{equation}
Equation (\ref{field_eq_k0}) showcases the breaking of statistical isotropy when $\Sigma_1\ne\Sigma_2$. It also indicates that SOI 
naturally imprint \textit{birefringence} 
to the random medium. From Eq. (\ref{field_eq_k0}), we infer 
\begin{equation}
\boldsymbol{e}(z)=\frac{\hat{\boldsymbol{x}}'+e^{i\phi+iz/2z_\text{L}}e^{-z/2z_\text{S}}\hat{\boldsymbol{y}}}{\sqrt{1+e^{-z/z_\text{S}}}},
\label{polar_evolution}
\end{equation}
whose evolution is represented on the Poincar\'{e} sphere of Fig. \ref{fig-sphere} for a circularly polarized beam, in the three configurations discussed above.
In case A, the polarization directly turns from circular to linear following the shortest path on the Poincar\'e sphere. In contrast, in configuration B the oscillating SHE is associated with a permanent, periodic evolution of the helicity. In the generic case C, finally, the oscillation of helicity is damped and $\boldsymbol{e}(z\!\gg\! z_\text{S})$ converges to the attractor point $\hat{\boldsymbol{x}}'$.
Fundamentally, the dual evolutions of $R_y(z)$ and $\boldsymbol{e}(z)$ originate from the conservation of angular momentum
\begin{equation}
k_0 R_y(z)+V(z)=\text{constant}= V(0)=\sin\phi.
\label{conservation_eq}
\end{equation}
$V(z)=2\!\int\! d\br_\perp\Im(\langle E^*_x \rangle\langle E_y \rangle)/\!\int d\br_\perp I_c$ is the spin angular momentum (i.e., the amount of circularly polarized light in the beam, $V$ axis in Fig. \ref{fig-sphere}) and $k_0 R_y(z)$ is the beam  orbital momentum, with Eq. (\ref{conservation_eq}) implying a mutual conversion between the two \cite{Andrews13}.
For $\phi=\pi/2$ and in case A, the initial spin is monotonically converted into orbital momentum while the beam is monotonically shifted. 
In cases B and C in contrast, the exchange between $V$ and $R_y$ involves successive mutual conversions, hence the periodic trajectories.
Together with Eq. (\ref{field_eq_k0}), the conservation of angular momentum
also sheds light on the second term in Eq. (\ref{beam_shift_eq}), which predicts a SHE for a spinless incident beam, $V(0)=0$.
In that case, the beam \textit{spontaneously} acquires a finite spin thanks to the birefringence  effect, Eq. (\ref{field_eq_k0}). A transverse motion then spontaneously appears so to satisfy Eq. (\ref{conservation_eq}). 
Interestingly, the SOI birefringence resembles the magneto-optic Voigt  (or Cotton-Mouton) effect, where a magnetic field perpendicular to the direction of propagation converts a linear polarization into an elliptic one \cite{Budker02}. In our scenario, the role of the magnetic field is played by $\bk_0$. 
The analogy ends here though, since no transverse motion arises in the Voigt effect.
\begin{figure}
	\centering{}\includegraphics[width=1\linewidth]{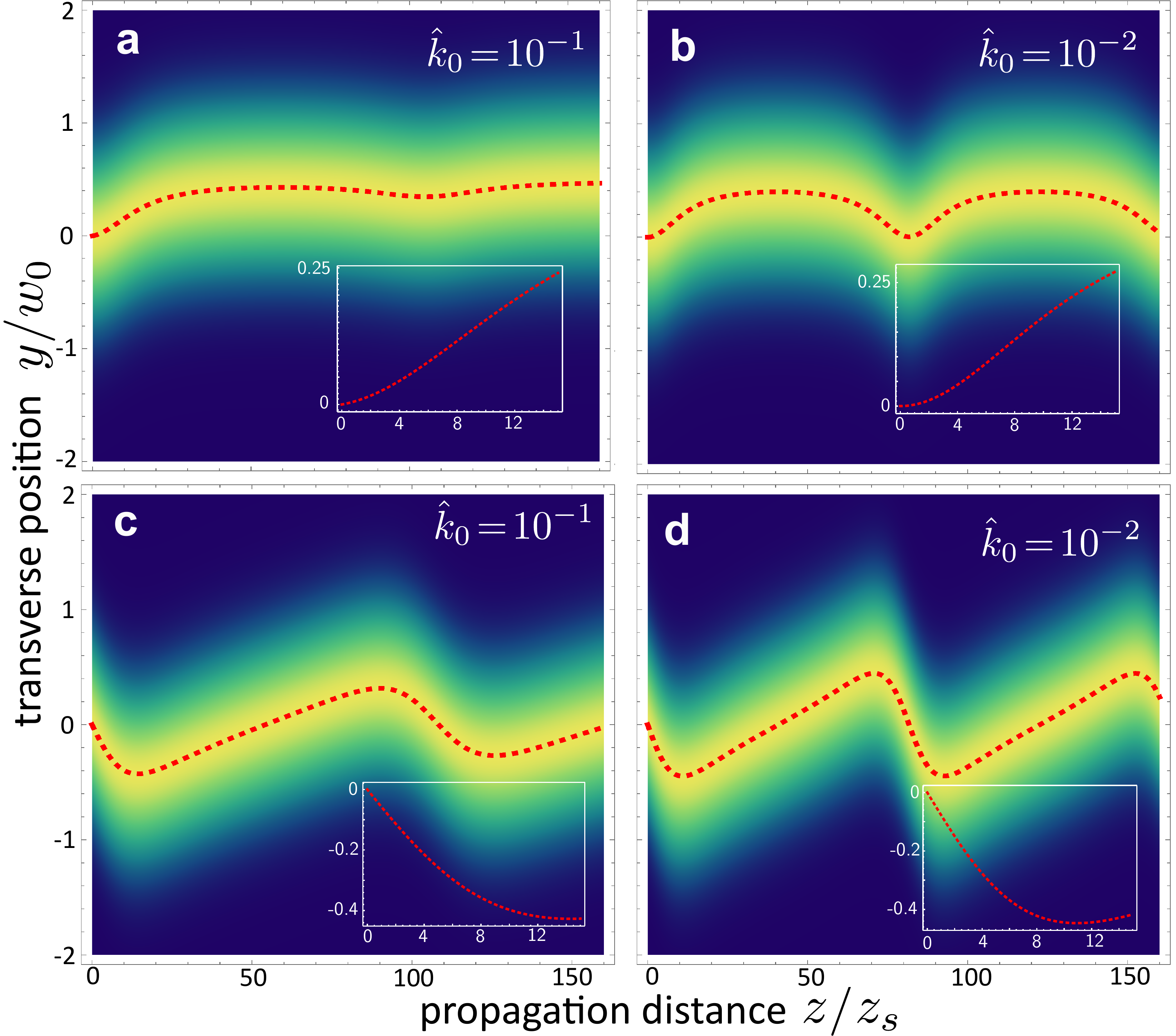}
 \caption{
Normalized intensity
 of the coherent mode post-selected by a polarizer $\boldsymbol{e}_\text{out}$ (``weak measurement''): $|\langle\boldsymbol{e}^*_\text{out}\cdot \bs{E}(\br_\perp,z)\rangle|^2/\int d\br_\perp |\langle\boldsymbol{e}^*_\text{out}\cdot \bs{E}(\br_\perp,z)\rangle|^2$. 
In panels a and b, $\boldsymbol{e}_\text{out}\propto\hat{\boldsymbol{y}}+ i \delta \hat{\boldsymbol{x}}'$, while in panels c and d, $\boldsymbol{e}_\text{out}\propto\hat{\boldsymbol{y}}+ \delta \hat{\boldsymbol{x}}'$, corresponding to an amplification of the first and second term in Eq. (\ref{beam_shift_eq}), respectively [here $k_0w_0=15$ and $\delta=(k_0w_0)^{-1}$]. Notice that the SHE is now of the order of the beam width $w_0$. Dotted curves mark the  centroid position $R_y(z)$.
Here, $k_0\sigma$ is chosen at the point where the ratio $z_s/z_\text{L}$ is maximum (see inset of Fig. \ref{zL_zS_plots}): $k_0\sigma=0.1056$ in panels a and c, $k_0\sigma=0.01005$ in panels b and d. 
Insets shows a zoom of the $R_y(z)$ at the scale of a few $z_s$.
}
 \label{fig_trajectories}
\end{figure}

Under normal conditions, the oscillating SHE discovered here is tenuous for two main reasons. First, because it occurs at the scale $k_0^{-1}$. Although this scale greatly exceeds the  optical wavelength, it remains small compared to the beam width $w_0$. Second, because the coherent mode attenuates as $\exp(-z/z_s)$ due to multiple scattering.
The first difficulty can be circumvented by exploiting 
the principle of weak quantum measurements \cite{Dressel14, Hosten08, Gorodetski12, Dennis12}. The idea is to start from a beam linearly polarized along $\boldsymbol{e}_0=\hat{\boldsymbol{x}}'$. SOI then split the beam into two parts, whose far tails have finite and opposite helicities. 
By detecting light using a nearly-orthogonal polarizer $\boldsymbol{e}_\text{out}\propto\hat{\boldsymbol{y}}+ i \delta \hat{\boldsymbol{x}}'$ with $|\delta |\ll1$, one can then select out these tails and amplify the first term in  Eq. (\ref{beam_shift_eq}) from $k_0^{-1}$ to $w_0$. In the SM we show that a similar strategy can be used to enhance the second term in Eq. (\ref{beam_shift_eq}), using a post-selection polarizer $\boldsymbol{e}_\text{out}\propto\hat{\boldsymbol{y}}{+} \delta \hat{\boldsymbol{x}}'$. 
To overcome the attenuation of the coherent mode due to scattering, on the other hand, one can again take advantage of the disorder correlation. 
Indeed, as shown in the inset of Fig. \ref{zL_zS_plots}, at a given angle $\hat{k}_0$ the ratio $z_s/z_\text{L}$ reaches a maximum at a specific $\sigma$, where $z_\text{L}$ reduces to a few $z_s$.
In Fig. \ref{fig_trajectories}, we show several types of transverse trajectories of the coherent mode computed by combining the weak measurement technique together with the maximization of $z_s/z_\text{L}$ via $\sigma$. The insets demonstrate the possibility to realize a sizable $R_y(z)$ over distances $z/z_s$ where the coherent mode remains observable.

We have unveiled the existence of oscillating spin Hall effects in transversally random media, controllable and amplifiable with the disorder correlation. 
Generally speaking, the control of SOI in disorder could be further improved using other degrees of freedom such as scatterer resonances \cite{Lagendijk96} or time-dependent beams \cite{Chalony11, Kwong14}. Spatio-temporal SOI seem, in particular, a promising direction of research \cite{Hancock19, Chong20, Bliokh21}. 

The authors thank D. Delande for fruitful discussions. This project has received financial support from the CNRS through the 80'Prime  program, and from the Agence Nationale de la Recherche (grant ANR-19-CE30-0028-01 CONFOCAL).


\begin{thebibliography}{99}
\bibitem{footnote1}
Eq. (\ref{beam_shift_eq}) holds at leading order in $\smash{\hat{k}_0^2}$. At arbitrary value of $\smash{\hat{k}}_0$, the expression remains correct provided the $1/k_0$ prefactors are replaced by $\smash{(1-\hat{k}_0^2)^{1/2}/k_0}$. This implies that SOI also show up at large angle of incidence, although their amplitude tends to decrease when approaching the regime of grazing incidence.
\end{thebibliography}
\end{document}